\documentclass[twocolumn,showpacs,preprintnumbers,amsmath,amssymb,prb,citeautoscript]{revtex4}

\usepackage{graphicx}
\usepackage{dcolumn}
\usepackage{bm}

\begin{document}


\title{Measurement of phosphorus segregation in silicon at the atomic-scale using STM}

\author{Lars Oberbeck, Neil J. Curson, Toby Hallam, Michelle Y. Simmons, and Robert G. Clark}

\affiliation{%
Centre for Quantum Computer Technology\\School of Physics\\The
University of New
South Wales\\Sydney NSW 2052, Australia}%

\author{Gerhard Bilger}%
\affiliation{%
Institute of Physical Electronics\\
University of Stuttgart\\
Pfaffenwaldring 47\\
70569 Stuttgart, Germany\\
}%

\date{\today}

\begin{abstract}
In order to fabricate precise atomic-scale devices in silicon
using a combination of scanning tunnelling microscopy (STM) and
molecular beam epitaxy it is necessary to minimize the
segregation/diffusion of dopant atoms during silicon
encapsulation. We characterize the surface segregation/diffusion
of phosphorus atoms from a $\delta$-doped layer in silicon after
encapsulation at 250$^{\circ}$C and room temperature using
secondary ion mass spectrometry (SIMS), Auger electron
spectroscopy (AES), and STM. We show that the surface phosphorus
density can be reduced to a few percent of the initial
$\delta$-doped density if the phosphorus atoms are encapsulated
with 5 or 10 monolayers of epitaxial silicon at room temperature.
We highlight the limitations of SIMS and AES to determine
phosphorus segregation at the atomic-scale and the advantage of
using STM directly.
\end{abstract}

\maketitle

\section{\label{sec:level1}Introduction}

If electronic devices continue to shrink at the same rate as
predicted by Moore's law \cite{Moore65} they will within the next
10 - 15 years reach the atomic-scale. The fabrication of
silicon-based devices at the atomic-scale requires the ability to
control the position of dopant atoms, such as phosphorus with
atomic precision. Dopant atoms are usually introduced into the
silicon crystal either by ion implantation or by diffusion from
highly doped surface layers \cite{Sze85}. However, these
techniques do not allow for atomically precise control of the
position of dopant atoms. An alternative way is to dope a 2D
silicon surface and subsequently overgrow the sheet of dopant
atoms with epitaxial silicon
\cite{Oberbeck_APL02,O'Brien_PRB01,Gossmann85}. One concern with
this method is that dopant atoms are known to diffuse at elevated
temperature \cite{Fahey89} and to segregate during epitaxial
silicon growth \cite{Nuetzel96}. It is therefore important to
monitor the precise position of dopant atoms after high
temperature processes such as crystal growth or annealing to
ensure segregation and diffusion are kept to a minimum.

Secondary ion mass spectrometry (SIMS) is a commonly used
measurement technique to investigate dopant segregation in Si
\cite{Nuetzel96,Friess92,Kasper00}. However, it only has the depth
resolution of a few nanometers depending on the sample and the
sputtering conditions \cite{Wilson89}. In the case of phosphorus
in Si, additional problems arise with SIMS measurements due to the
high mass resolution required to distinguish between $^{31}$P
(mass 30.9738 \cite{webelements}) and $^{30}$SiH (mass 30.9816
\cite{webelements}). Finally, ion beam mixing during the
sputtering process can broaden the dopant profiles thereby
influencing any quantitative analysis of dopant segregation.

In this paper, we compare three different methods to quantify the
segregation of P atoms in silicon. We use SIMS to study
segregation of P atoms in Si at the nanometer scale, Auger
electron spectroscopy (AES) for studies at the sub-nanometer
scale, and scanning tunnelling microscopy (STM) to measure the
segregation and diffusion of phosphorus atoms to the Si surface at
the atomic-scale. In each case a Si(001) surface is
${\delta}$-doped using PH$_3$ gas as the dopant source followed by
an anneal at 600$^{\circ}$C to incorporate the P atoms into the
surface. In order to determine the amount of segregation that can
occur during Si encapsulation the sheet of dopants is then
overgrown with several monolayers (ML) of epitaxial silicon either
at room temperature (RT) or 250$^{\circ}$C.

We demonstrate that reducing the silicon growth temperature to
room temperature can minimize the segregation length of P in Si to
sub-nanometer values, the lowest segregation length values for P
in Si shown so far. Subsequent sample annealing, necessary to
enhance the structural quality of the low temperature grown
epitaxial Si layer, causes the encapsulated P atoms to diffuse to
the surface. However, the density of P atoms that reach the
surface can be reduced to a few percent of the original P atom
density (even after annealing at 500$^{\circ}$C) if silicon
encapsulation is performed at RT.

\section{\label{sec:level1}Experimental}

Experiments were performed using an Omicron variable temperature
STM with Auger electron spectroscopy and a Si sublimation cell.
The base pressure of the two-chamber analysis/growth ultra-high
vacuum (UHV) system was ${<}$ 5 ${\times}$ 10$^{-11}$ mbar.
Phosphorus doped 1 ${\Omega}$cm Si(001) wafers were cleaved into 2
${\times}$ 10 mm$^{2}$ sized samples, mounted in sample holders,
and then transferred into the UHV system. Sample preparation was
performed in vacuum without prior ex-situ treatment by outgassing
overnight at ${\sim}$550$^{\circ}$C using a resistive heater
element followed by direct current heating at the same temperature
for several hours. The sample was then flashed for 30 to 60 s to
1150$^{\circ}$C by passing a direct current through the sample.
After flashing, the sample was cooled slowly
(${\sim}$3$^{\circ}$C/s) from 900$^{\circ}$C to room temperature.
The clean sample surface at room temperature was then directly
exposed for 15 min to a beam of phosphine (PH$_{3}$) molecules at
a chamber pressure of 10$^{-9}$ mbar resulting in a surface
saturation dosing with phosphine molecules of 0.37 ML
\cite{Lin99}. An annealing step at 600$^{\circ}$C for 1 min
incorporated P atoms into the Si surface and desorbed the H atoms
resulting from dissociation of the PH$_{3}$ molecules giving a
total density of 0.25 ML of P atoms at the surface \cite{Lin99}.

For the fabrication of $\delta$-doped samples for SIMS
measurements, the sheet of P atoms was then overgrown by an
$\sim$24 nm thick epitaxial intrinsic Si (${\rho}$ ${>}$ 1
k${\Omega}$cm) layer deposited from a Si sublimation cell at a
substrate temperature of either 250$^{\circ}$C or RT, and a growth
rate of 2 ML/min. Note that RT growth means without intentional
sample heating. However, during deposition the sample was
radiatively heated by the glowing Si arc used as a sublimation
source for Si deposition. The sample temperature is estimated to
increase to a few 10$^{\circ}$C above RT during Si deposition. For
growth at 250$^{\circ}$C the Si sample was heated by direct
current flow through the sample and the sample temperature was
controlled by an infrared pyrometer. SIMS measurements were
performed on the ${\delta}$-doped samples to determine the level
of surface segregation of the P atoms during epitaxial silicon
overgrowth. Using an ATOMIKA 6500 system with Cs$^+$ primary ion
beam the P depth profile was recorded at primary ion energies of
2, 5.5, and 15 keV. As the mass resolution of the ATOMIKA system
was not sufficient to distinguish between the masses of $^{31}$P
and $^{30}$SiH, additional SIMS data was acquired at 14.5 keV
using a CAMECA SIMS system with a higher mass resolution. The SIMS
results were then used to quantify the segregation of phosphorus
atoms at the nanometer scale.

In addition, Auger electron spectroscopy and STM were used to
study the segregation and diffusion of P atoms at the
sub-nanometer scale and atomic-scale, respectively, during
epitaxial Si growth and subsequent annealing steps. As before, a
Si(001) surface was phosphorus $\delta$-doped and then overgrown
by several monolayers of intrinsic Si at a growth rate of 2 ML/min
and at substrate temperatures of either $\sim$250$^{\circ}$C or
room temperature. After growth the sample was transferred to the
analysis chamber where the Auger electron spectroscopy and STM
measurements of the density of P atoms segregated to the Si
surface were carried out. For Auger electron spectroscopy
measurements the sample was then annealed in successive steps at
temperatures between $\sim$300 and 900$^{\circ}$C in
150$^{\circ}$C steps for 1 min at each step. Auger electron
spectroscopy measurements were then performed after cooling down
of the sample to RT. For separate STM experiments sample annealing
was carried out between 350 and 950$^{\circ}$C in
${\sim}$50$^{\circ}$C steps for 5 s at each step. After each
process step the sample surface was investigated at room
temperature and the number of P atoms at the surface counted.

\section{\label{sec:level1}Results and Discussion}

\subsection{\label{sec:level2}SIMS measurements to study segregation at the nanometer scale}
Figure 1 (a) shows SIMS measurements of phosphorus
${\delta}$-doped Si(001) samples overgrown at 250$^{\circ}$C and
RT, respectively. Using the ATOMIKA system, two distinct mass-31
peaks in the two curves at a depth of ${\sim}$10 and ${\sim}$24 nm
were measured. Since it is difficult to distinguish between the
$^{30}$SiH and $^{31}$P peaks additional measurements were carried
out using a high mass resolution CAMECA system which displayed
only one $^{31}$P peak at ${\sim}$ 20 nm (see inset of Fig. 1
(a)). This additional SIMS data confirms that the broad peak at
${\sim}$10 nm is not due to $^{31}$P, but is $^{30}$SiH arising
from H adsorption at the Si surface. The peak at 24 nm can
however, be ascribed to the P ${\delta}$-doped layer at the
interface between the epitaxial layer and the substrate. The
$^{31}$P peaks of the samples grown at 250$^{\circ}$C and RT,
respectively, display a full width at half maximum (FWHM) of 6 and
4 nm, respectively, for a primary ion energy of 5.5 keV
demonstrating the smaller amount of P segregation occurring during
RT Si overgrowth. The segregation length \cite{Nuetzel96}
$\triangle$ for P segregation in Si was obtained by fitting the
exponential decrease of the P concentration towards the sample
surface displayed in Fig. 1 (a), giving values of $\triangle$ =
2.3 nm and 1.5 nm for the samples grown at 250$^{\circ}$C and RT,
respectively.

Figure 1 (b) highlights the limitations of the SIMS technique.
Here we present SIMS measurements of the phosphorus
${\delta}$-doped Si(001) sample overgrown at 250$^{\circ}$C for
various Cs$^+$ primary ion energies of 2, 5.5, and 15 keV. Note
that the apparent reduced P peak height of the 2 keV data results
from a quantification of the data using the 5 keV signal as a
reference. The P peak is observed to decay into the substrate with
the decay length increasing from ${\sim}$1 decade/6.5 nm to
${\sim}$1 decade/20 nm as the primary ion energy is increased from
2 to 15 keV. This variation of the decay length with ion energy is
a well-known measurement artefact arising from ion beam mixing,
i.e. primary Cs$^+$ ions pushing P atoms further into the
substrate during the sputtering process. The SIMS data of the
${\delta}$-doped sample shows a FWHM of the $^{31}$P peak of 5, 6,
and 14 nm for Cs$^+$ primary ion energies of 2, 5.5, and 15 keV,
respectively. While it can be seen that measurement artefacts due
to ion beam mixing decrease with decreasing primary ion energy, it
is also obvious that a small primary ion energy of 2 keV results
in low count rates and therefore a stronger noise in the
measurement signal and a higher detection limit. Most importantly
the variation of the shape of the P peak with primary ion energy
demonstrates that the resolution of the SIMS technique is only of
the order of a few nm. As a result the segregation of P atoms
during growth can only be detected at this level using SIMS
highlighting that the technique is not sensitive enough to measure
the segregation at the atomic-scale.

\begin{figure}
\includegraphics{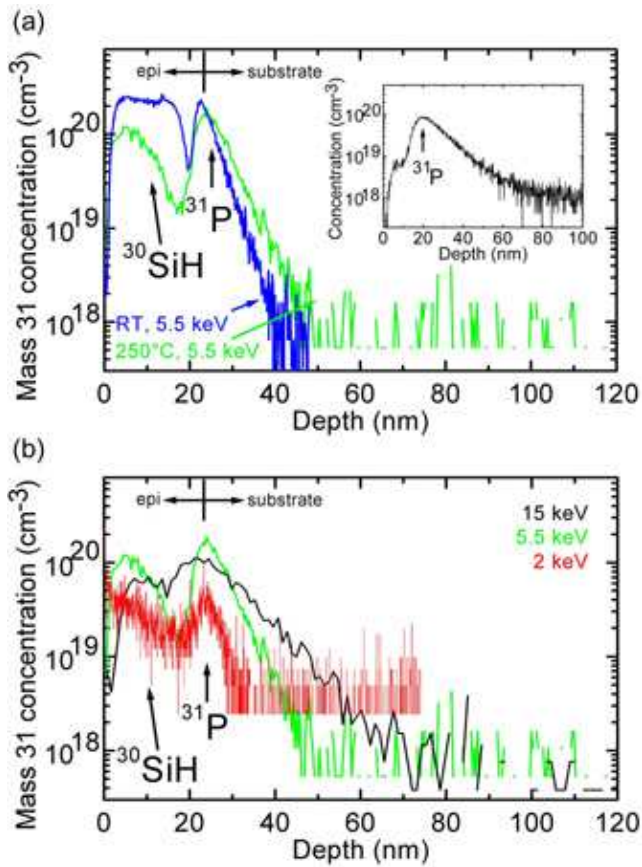}
\caption{\label{fig:one} (a) Mass 31 depth profile of
${\delta}$-doped samples grown at 250$^{\circ}$C and room
temperature, respectively, determined by SIMS using a 5.5 keV
Cs$^{+}$ primary ion energy in an ATOMIKA system. The inset shows
the high mass resolution CAMECA SIMS measurement. (b) Mass 31
depth profile of the ${\delta}$-doped sample grown at
250$^{\circ}$C using 2, 5.5, and 15 keV Cs$^{+}$ primary ion
energies in an ATOMIKA system.}
\end{figure}

\subsection{\label{sec:level2}AES measurements to study segregation at the sub-nanometer scale}
As a consequence we also carried out studies of phosphorus
segregation and diffusion at the sub-nanometer scale using Auger
electron spectroscopy. Fig. 2 (a) shows Si and P signals at 107 eV
and 120 eV Auger electron energies respectively, in the
differentiated AES spectrum of the Si sample surface obtained from
(i) the clean Si surface, the surface after (ii) PH$_3$ saturation
dosing and P incorporation, (iii) 5 ML Si overgrowth at RT, and
(iv) various 1 min annealing steps at the temperatures displayed.
Note: the intensity of the Auger signal decreases with sample
depth and is therefore limited to approx. the first three ML in Si
\cite{Seah79}.

In order to quantify the amount of P segregation/diffusion we have
analyzed the AES data of the integrated spectrum. The number of
Auger electrons calculated from the integrated spectrum is
directly proportional to the P surface density \cite{Jacobson98}.
Fig. 2 (b) shows the P density in the first three ML of the sample
relative to the P density of the saturation dosed surface from two
different growth experiments at 250$^{\circ}$C and RT. If we
consider the figure in more detail we see that the clean Si
surface after flashing does not show any discernable P
concentration, as expected. After P saturation dosing and
incorporation of the P atoms, the maximum P density normalized to
1 is visible on the surface giving the largest Auger electron peak
at 120 eV. The P surface density then decreases to $\sim$0.74 and
$\sim$0.12 after encapsulation with 5 ML of Si grown at
250$^{\circ}$C and RT, respectively, and increases again with
sample annealing at various temperatures up to 600$^{\circ}$C due
to P diffusion to the surface. After annealing at 750$^{\circ}$C
the P density decreases markedly due to P desorption from the
surface as P$_2$ \cite{Jacobson98}. Finally after annealing to
900$^{\circ}$C the P peak disappears completely. Phosphorus is
still detectable after the 750$^{\circ}$C annealing step as P
diffusion from subsurface layers to the surface is still occurring
during the anneal.

To calculate the segregation length $\triangle$ of P in Si at
250$^{\circ}$C and RT from the AES data we use the relation
\cite{Nuetzel96}

\begin{equation}\label{eq:one}
p_{inc} = \emph{a}_0/4\triangle
\end{equation}

with the incorporation probability $p_{inc}$ and $\emph{a}_0/4$ =
0.1358 nm the distance between two subsequent Si(001) monolayers.
The relation between the incorporation probability and the
segregation length is only valid for first-order kinetic processes
and assumes negligible desorption and bulk diffusion. To analyze
our data we know that in Fig. 2 (b) we find $\sim$74$\%$ of P
atoms at the surface after 5 ML Si growth at 250$^{\circ}$C. From
this value we get an incorporation probability $p_{inc}$ = 0.06.
According to equation 1 this value corresponds to a segregation
length of $\triangle$ = 2.3 nm. For RT growth we find a value of
$\triangle$ = 0.39 nm. In section D we discuss these values of
segregation length obtained using AES measurements in relation to
both the values in the literature and to segregation lengths
obtained from SIMS and STM measurements. We note however that AES
is sensitive to subsurface, non-segregated P atoms and therefore
we expect these values to be slightly on the high side of the real
$\triangle$. In addition, in the case of RT Si growth due to the
high surface roughness of the as-grown surface parts of the
surface are covered by less than 5 ML of Si which makes it
possible that AES detects non-segregated P atoms in their original
lattice positions.

\begin{figure}
\includegraphics{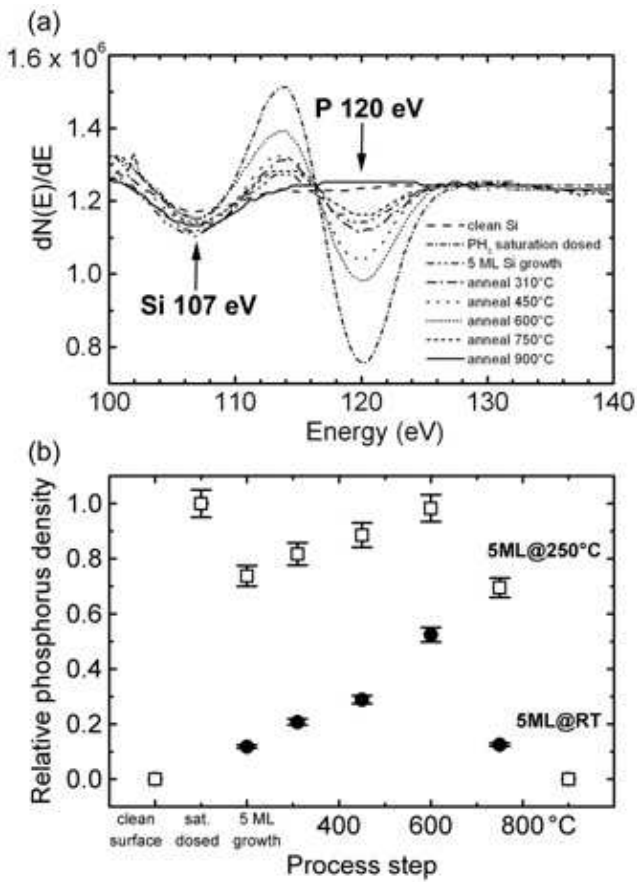}
\caption{\label{fig:two} (a) Differentiated Auger electron
spectroscopy spectrum of the clean Si surface and the surface
after PH$_3$ saturation dosing and annealing, 5 ML Si growth at
room temperature, and various annealing steps for 1 min each. (b)
Phosphorus density determined from integrated Auger electron
spectra normalized to the phosphorus density of the saturation
dosed surface for Si encapsulation at 250$^{\circ}$C and RT.}
\end{figure}

\subsection{\label{sec:level2}STM measurements to study segregation at the atomic-scale}
Finally in order to gain a better quantitative estimate of the P
segregation/diffusion at the atomic-scale we used the STM
directly. We imaged the clean Si(001) surface, the surface after
PH$_3$ dosing/P incorporation and after Si growth and subsequent
annealing steps to determine the density of P atoms at the
surface. Annealing of the sample after growth at 255$^{\circ}$C or
RT is necessary to flatten the surface in order to identify the
characteristic asymmetric appearance of the Si-P heterodimers
which form after P incorporation from adsorbed PH$_3$ molecules
\cite{Oberbeck_APL02}.

Figures 3 (a) - (f) show filled state STM images taken at room
temperature of two separate growth experiments. For each
experiment STM images of the clean Si(001) surface are shown after
initial sample flashing (a); P incorporation (b); Si overgrowth at
different temperatures (c); and successive annealing steps for 5 s
at ${\sim}$350$^{\circ}$C, ${\sim}$500$^{\circ}$C, and
${\sim}$600$^{\circ}$C (d) - (f). Note that for space reasons only
selected STM images of the annealing sequence are shown here. The
complete annealing process is however summarized in Fig. 5 (a) and
(b). The two experiments only differ in the Si overgrowth step
(Fig. 3 (c)): either 5 ML of Si were grown at 255$^{\circ}$C or at
RT.

Both experiments started with the preparation of a low defect
density clean Si(001) surface, see Fig. 3 (a). The surface was
then phosphine dosed and annealed at 600$^{\circ}$C for 1 min to
incorporate the phosphorus atoms into the surface and desorb the
associated hydrogen \cite{Pietsch95}. The resulting image in Fig.
3 (b) shows dimer rows on the Si surface that are strongly buckled
due to the high phosphorus density at the surface. After 5 ML
growth at 255$^{\circ}$C the surface shows small 2D islands and
short Si dimer chains (c). Subsequent annealing of the surface at
temperatures of 345 (d), 498 (e), and 600$^{\circ}$C (f) causes
the Si surface to flatten due to island coarsening and diffusion
of Si atoms to step edges. Simultaneously, the density of bright
asymmetric features that result from segregated P atoms forming
Si-P heterodimers at the surface increases in accordance with
Auger electron spectroscopy data in Fig. 2 (b).

The difference between the two experiments is in the Si growth
step (Fig. 3 (c)), where at RT 3D Si islands are formed due to the
small mobility of Si atoms on the surface \cite{Herman96}.
Successive annealing steps flatten the surface as with the 5 ML
growth at 255$^{\circ}$C. Interestingly, the Si surface morphology
in the different experiments looks very similar for the same
annealing temperature, even for the first annealing step at
$\sim$350$^{\circ}$C for 5 s (d). However, Figs. 3 (d) - (f)
clearly show that the density of Si-P heterodimers at the Si
surface is much lower for a given anneal temperature if the Si
encapsulation occurs at RT compared with 255$^{\circ}$C Si growth.

\begin{figure}
\includegraphics{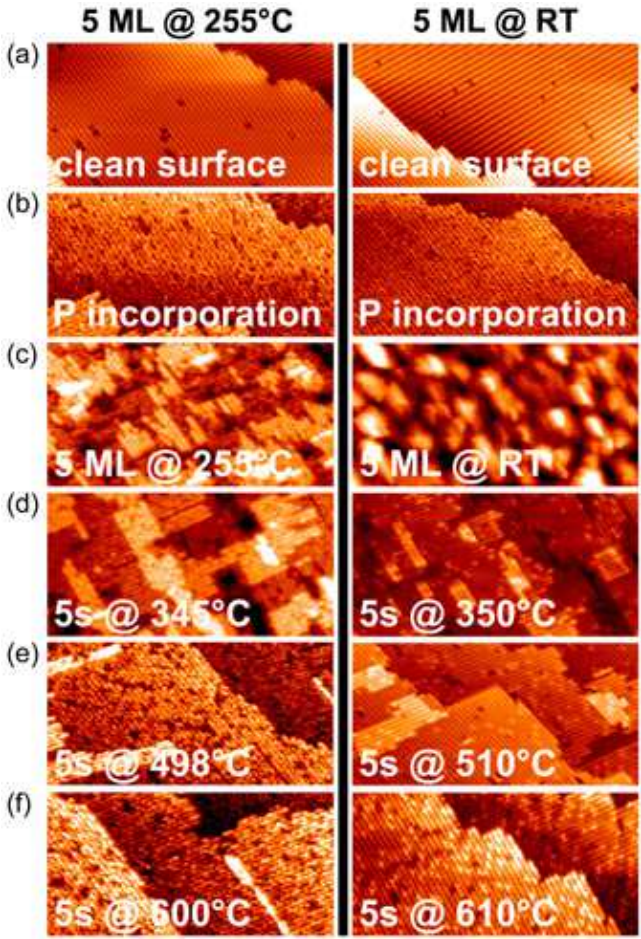}
\caption{\label{fig:three} Filled state STM images showing the
clean Si(001) surface (a) and the surface after phosphorus
incorporation (b)/Si overgrowth (c)/annealing at various
temperatures (d) - (f) from two experiments: 5 ML Si growth at
255$^{\circ}$C (left column) and room temperature (right column).
STM images taken after various annealing steps show an increasing
density of bright features at the surface which are Si-P
heterodimers. Image size of individual STM images is 50 ${\times}$
25 nm$^2$.}
\end{figure}

Figure 3 (c) highlights an important limitation in the use of STM
to investigate P segregation/diffusion: the high surface roughness
that results from low temperature growth, especially room
temperature growth, makes identification of Si-P heterodimers at
the surface difficult. This occurs for two reasons: firstly, STM
images of rough Si surfaces usually have a lower quality compared
to images of an atomically flat surface. Secondly, to identify a
Si-P heterodimer a sufficient brightness contrast between the
heterodimer and the surrounding Si surface is necessary, which can
only be obtained for atomically flat surfaces. Therefore, Si-P
heterodimers only become clearly visible in our experiments after
a short anneal at $\sim$350$^{\circ}$C for 5 s. As a consequence
the P density at the Si surface after the first anneal is a result
of not only P segregation that occurs during growth but also
arises from diffusion of P atoms that occurs during the subsequent
anneal. Nonetheless it is reasonable to assume that P diffusion,
with an energy barrier of 3.66 eV \cite{Fahey89}, is negligible
for short anneals at such low temperatures. Thus, the P density
observed at the surface after the first anneal is most likely a
result of P segregation that occurs during Si growth.

To further minimize the surface P density we here also
encapsulated the sheet of P atoms with 10 ML, instead of only 5
ML, of Si grown at RT. Figures 4 (a) - (d) show a comparison
between the two experiments. For both experiments the surface
after Si growth consists of small 3D islands and flattens during
subsequent annealing steps as before. As expected, the STM images
show that the density of Si-P heterodimers at the Si surface is
higher for 5 ML encapsulation compared to 10 ML Si growth.

\begin{figure}
\includegraphics{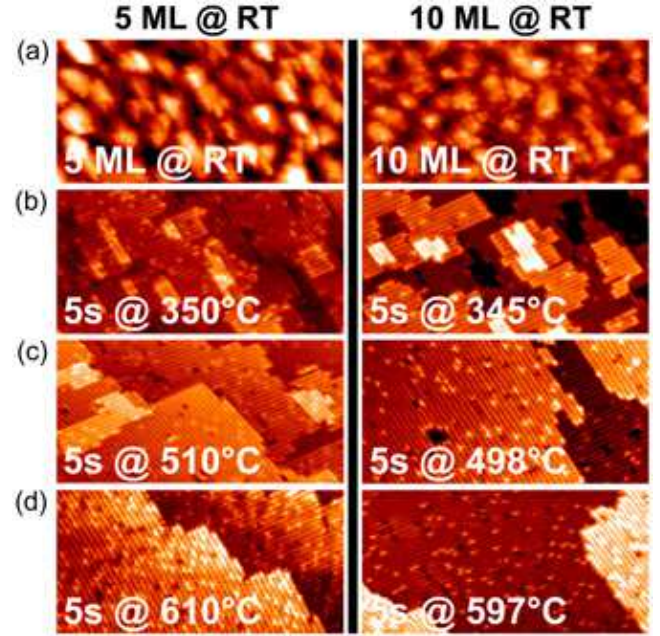}
\caption{\label{fig:four} Filled state STM images showing the
Si(001) surface after Si overgrowth (a) and annealing at various
temperatures (b) - (d) from two experiments: 5 ML (left column)
and 10 ML Si growth at RT (right column). Image size of individual
STM images is 50 ${\times}$ 25 nm$^2$.}
\end{figure}

To quantify the density of P atoms at the Si(001) surface observed
in the STM images of Figs. 3 and 4 we have counted the number of
Si-P heterodimers after each anneal. Figure 5 (a) shows the
increase in the density of P atoms at the surface following
subsequent annealing steps for the three different experiments.
The relative density is obtained by comparison of the phosphorus
density after Si growth and sample annealing, with the initial
density of incorporated phosphorus atoms after phosphine dosing of
the clean Si surface.

If the encapsulation of the sheet of P atoms is undertaken at
255$^{\circ}$C, already ${\sim}$25\% of the initial number of
phosphorus atoms have segregated during Si growth to occupy
surface lattice sites after the first 5 s annealing step at the
lowest applied annealing temperature (${\sim}$350$^{\circ}$C), see
Fig. 5 (a). After subsequent annealing at 400, 450, 500, 550, to
600$^{\circ}$C nearly 60\% of the phosphorus atoms are present at
the surface. These results demonstrate that even with a short, low
temperature anneal encapsulation of P atoms in epitaxial Si grown
at $\sim$250$^{\circ}$C results in significant P segregation.

In contrast, if P atoms are overgrown with 5 ML of Si deposited at
RT and annealed at 350$^{\circ}$C for 5 s (Fig. 5 (a)), only
$\sim$5\% of the initial number of P atoms are present at the
surface (a factor of 5 less). During subsequent annealing at 400,
450, 500, 550, and 600$^{\circ}$C the P density only increases
slightly to $\sim$10\% due to diffusion of phosphorus atoms from
subsurface layers to the surface. The reduced density of P atoms
at the Si surface compared to the 255$^{\circ}$C Si growth
experiment is a direct consequence of the strongly suppressed
segregation of P atoms during Si overgrowth at RT. Figure 5 (a)
also shows that the surface P density can be further reduced by
growth of a thicker Si layer of 10 ML at RT, as seen in the STM
images of Figs. 4 (b) - (d). Note that the relative P
concentration after various annealing steps obtained from Auger
electron spectroscopy measurements shown in Fig. 2 (b) is higher
than the values displayed in Fig. 5 (a). This is a direct result
of the longer annealing time of 1 min for each step in the AES
experiments and the fact that AES also detects subsurface P atoms,
highlighting the limitations of this technique.

Figure 5 (b) shows the decrease of the surface phosphorus density
for the case of 5 ML Si growth at 255$^{\circ}$C after successive
anneals at temperatures between 600 and 950$^{\circ}$C due to
phosphorus desorption from the surface. This figure also
demonstrates that P atoms are still present at the surface even
after annealing for 5 s at 900$^{\circ}$C due to the continual
diffusion of P atoms from subsurface layers to the surface.

Using equation 1 we calculate the segregation length $\triangle$
of P in Si from the STM data. We know that in Fig. 5 (a) we find
about 25$\%$ of P atoms at the surface after 5 ML Si growth at
255$^{\circ}$C and the first annealing step at
$\sim$350$^{\circ}$C. This means that 75$\%$ of the P atoms were
incorporated. From this value (assuming that only segregation
during growth occurred and that the diffusion of P atoms during
the first annealing step at $\sim$350$^{\circ}$C is negligible) we
get an incorporation probability $p_{inc}$ = 0.24. According to
equation 1 this value corresponds to a segregation length of
$\triangle$ = 0.58 nm. For the 5 and 10 ML Si growth at room
temperature, the same analysis gives an average segregation length
of $\triangle$ = 0.34 nm.

\begin{figure}
\includegraphics{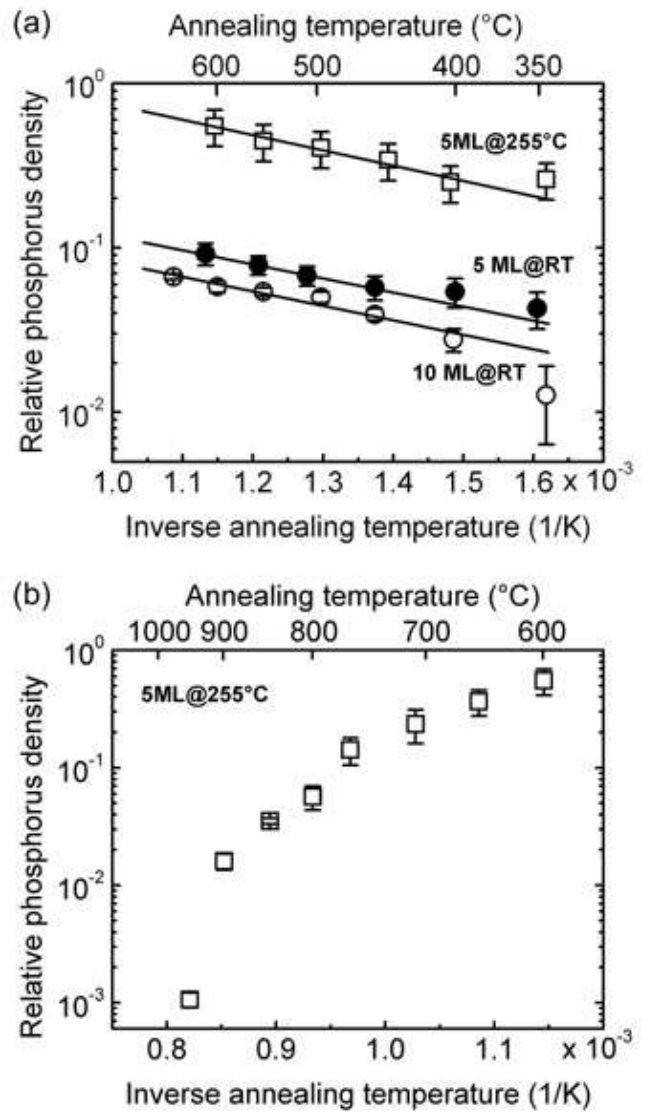}
\caption{\label{fig:five} Relative density of P atoms at the
Si(001) surface after Si encapsulation and various annealing
steps. The surface P density increases during annealing at
temperatures up to ${\sim}$600$^{\circ}$C (a) and decreases due to
P desorption from the surface at annealing temperatures $\gtrsim$
650$^{\circ}$C (b). The density relative to the initial coverage
was determined from STM images. The lines are guide to the eye.}
\end{figure}

\subsection{\label{sec:level2}Comparison of segregation length data}

In this section we compare the segregation length values obtained
from our SIMS, AES and STM data with values from the literature.
Segregation length values for P in Si have been published by the
Abstreiter group \cite{Friess92,Nuetzel96} and by Hobart \emph{et
al.} \cite{Hobart96} In these studies electrochemical
capacitance-voltage profiling (eCV), SIMS
\cite{Friess92,Nuetzel96}, and spreading resistance profiling
\cite{Hobart96} are used to determine the segregation length at
growth temperatures in the range of 320 to $\sim$900$^{\circ}$C
and growth rates in the range of 0.06 \AA/s to 2 \AA/s. To compare
the literature data with our experimental results we first
re-scale the literature data to the growth rate of 0.045 \AA/s = 2
ML/min used in our experiments using the relation \cite{Nuetzel96}

\begin{eqnarray}
\triangle \sim \sqrt{\frac{1}{R}}, \label{eq:two}
\end{eqnarray}

where $R$ is the Si growth rate. The growth rate dependence of the
segregation length arises from the fact that the growth rate
determines the time available for dopant atoms to exchange sites
from subsurface to surface lattice sites before they are
encapsulated in Si. Since the literature data was obtained at
higher Si growth temperatures we have extrapolated their data to
RT, as shown in Fig. 6.

From this figure we can see that \emph{three} observations can be
made. Firstly, the experimental values for P segregation in Si
determined by the Abstreiter group (Friess and N\"{u}tzel) and
Hobart \emph{et al.} differ significantly. This difference can
result for a number of different reasons, including a difference
in the temperature calibration of the growth systems, different
measurement techniques (eCV and SIMS vs spreading resistance
profiling), and different P doping profiles and doping techniques
(MBE growth of P doped layers vs P diffusion) used by the two
groups. As a consequence the extrapolation of experimental data
from Friess, N\"{u}tzel, and Hobart \emph{et al.} to lower
temperatures also results in a large difference in the low
temperature segregation lengths.

Secondly, there is a noticeable difference in the segregation
length values obtained from our SIMS, AES, and STM experiments at
the same temperature, see Table 1. The STM values are consistently
lower than the respective values determined from SIMS and AES
measurements. This can be attributed to limitations of these other
techniques, most notably: (i) The depth resolution of the SIMS
technique is limited to the nanometer range \cite{Wilson89} as a
result of the sputtering process, where P atoms in the sample are
measured from several monolayers simultaneously. This broadens the
P signal leading to an increase in the segregation length
obtained. (ii) AES not only detects segregated P atoms at the
surface but also detects subsurface, non-segregated P atoms. The
analysis of AES data therefore displays a segregation length which
is higher than the real $\triangle$. It is anticipated therefore
that the segregation lengths obtained from STM measurements are
more accurate than both SIMS/AES analysis.

\begin{table}
\caption{\label{tab:table1}Segregation length data obtained from
SIMS, AES, and STM measurements (values in nm). }
\begin{ruledtabular}
\begin{tabular}{lcr}
&RT&250$^{\circ}$C\\
\hline
SIMS & 1.50 & 2.30\\
AES & 0.39 & 2.30\\
STM & 0.34 & 0.58\\
\end{tabular}
\end{ruledtabular}
\end{table}

Thirdly, a final observation that can be made is that our
experimental data at these much lower temperatures lies at the
limits of the data range expected from extrapolation of the
literature values (see the grey area in Fig. 6). There could be
two main reasons for this. (i) The sample structures measured are
somewhat different: while we used P $\delta$-doped layers,
N\"{u}tzel \cite{Nuetzel96}, Friess \cite{Friess92} and Hobart
\cite{Hobart96} used bulk doped layers. (ii) The strong
temperature dependence of the segregation length can mean that a
small difference in the temperature calibration of the different
growth systems can give significant differences in the segregation
length.

Despite the small differences in the segregation length between
our data and extrapolated data from the literature, our study
shows that low temperature Si encapsulation of P atoms with a few
ML of Si, can suppress the segregation of P atoms very
effectively. Our results also demonstrate that STM, in contrast to
SIMS and AES, allows for a better quantitative determination of
the segregation length at the atomic-scale.

\begin{figure}
\includegraphics{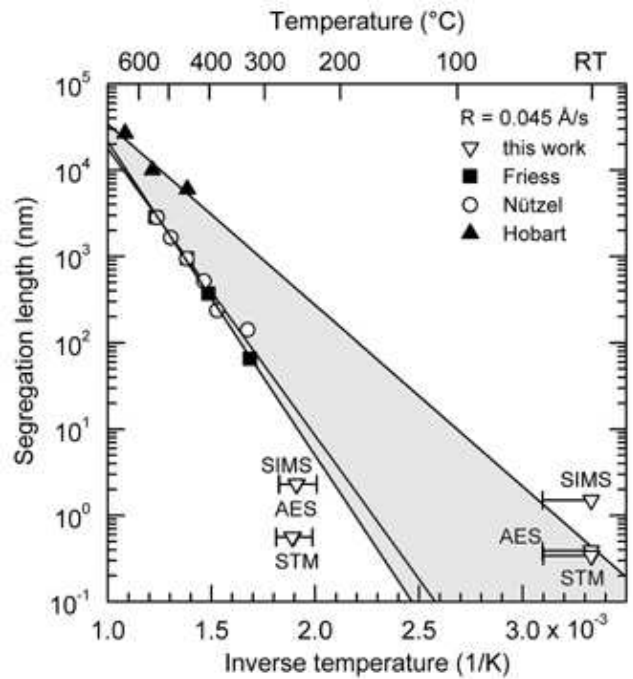}
\caption{\label{fig:six} Comparison of phosphorus segregation
length data with experimental data from the literature. Note that
all data were corrected for a growth rate $R$ = 0.045 \AA/s using
the relation $\triangle \sim 1/\surd{R}$ \cite{Nuetzel96}. Also
note that experimental data of Friess and N\"{u}tzel are from the
same group and differ significantly from Hobart's results. The
grey area denotes the range of values expected from literature
results.}
\end{figure}

Future experiments will focus on STM imaging of the buried
phosphorus dopant atoms once they are encapsulated by epitaxial Si
overgrowth. Buried dopant imaging has already been demonstrated
for boron and arsenic dopants in Si using hydrogen terminated
Si(001) surfaces \cite{Liu01}. Such a technique would be very
important to detect buried P atoms in the Si matrix, not only to
allow atomically precise determination of the lateral position,
but also the vertical position of the P atoms.

\section{\label{sec:level1}Conclusions}
We have used SIMS and, for the first time, AES and STM to analyze
P segregation/diffusion in Si at very low temperatures
(250$^{\circ}$C and room temperature). We show that SIMS is unable
to detect segregation lengths of phosphorus atoms in silicon below
the nanometer scale. In contrast, AES and STM can be used to
monitor the segregation and diffusion of phosphorus atoms in
silicon at the sub-nanometer and atomic-scale, respectively. We
highlight that Auger electron spectroscopy is limited as a
quantitative technique since it detects subsurface, non-segregated
P atoms. As a result we demonstrate that STM measurements can be
used directly to quantify the surface density of segregated P
atoms at the atomic-scale. Using STM we obtained segregation
lengths below 1 nm for phosphorus encapsulated in silicon grown at
250$^{\circ}$C and room temperature respectively, which are the
lowest values recorded in the literature. Differences between the
segregation length obtained and extrapolated experimental data
from the literature are accounted for by differences in sample
preparation and experimental techniques. These results highlight
the effectiveness of room temperature silicon encapsulation to
minimize dopant segregation in silicon and how it is possible to
quantify this directly using STM.

\begin{acknowledgments}
The authors thank P. Priestley for analysis of STM data. LO
acknowledges a Hewlett-Packard Fellowship. MYS acknowledges a
Federation Fellowship. This work was supported by the Australian
Research Council and the Army Research Office (ARO) under contract
number DAAD19-01-1-0653.
\end{acknowledgments}

\end{document}